\documentclass[11pt]{article}
\usepackage{moriond,epsfig}
\usepackage{amssymb}
\usepackage{hyperref}

\def\be{\begin{equation}}
\def\ee{\end{equation}}
\def\bea{\begin{eqnarray}}
\def\eea{\end{eqnarray}}

\newcommand{\dd}{{\rm d} }
\newcommand{\sqrts}{\sqrt{s}}

\def\pt{p_{_\perp}}
\def\xt{x_{_\perp}}
\def\X{{\rm X}}

\def\nactive{{n}_{\rm active}}

\def\nexp{n^{{\rm exp}}}
\def\nexpga{n^{{\rm exp}}_{{\gamma}}}
\def\nexpjets{n^{{\rm exp}}_{{\rm jets}}}
\def\nnlo{n^{{\rm NLO}}}

\def\thetacm{\vartheta}
\def\sigmainv{\sigma^{\rm inv}}

\def\Deltafit{{\Delta^{\rm fit}}}

\begin{document}
\vspace*{4cm}

\title{Higher-twist contributions to large $\pt$ hadron production in hadronic collisions}

\author{Fran\c{c}ois Arleo}
\address{LAPTH \footnote{Laboratoire d'Annecy-le-Vieux de Physique Th\'eorique, UMR5108, Universit\'e de Savoie, CNRS}, BP 110, 74941 Annecy-le-Vieux cedex, France}
\author{Stanley J. Brodsky}
\address{SLAC National Accelerator Laboratory, Stanford University, Stanford,
California 94309, USA\\
\& CP3-Origins,
University of Southern Denmark, Odense 5230 M, Denmark}
\author{Dae Sung Hwang}
\address{Department of Physics, Sejong University, Seoul 143--747, Korea}
\author{Anne M. Sickles}
\address{Brookhaven National Laboratory, Upton, New York 11973, USA}

\maketitle\abstracts{
The scaling behavior of large-$\pt$ hadron production in hadronic collisions is investigated. A significant deviation from the NLO QCD predictions is reported, especially at high values of $\xt~=~2\pt/\sqrt{s}$. In contrast, the prompt photon and jet production data prove in agreement with leading-twist expectations. These results are interpreted as coming from a non-negligible contribution of higher-twist processes, where the hadron is produced directly in the hard subprocess. Predictions for scaling exponents at RHIC are successfully compared to PHENIX preliminary measurements. We suggest to trigger on \emph{isolated} large-$\pt$ hadron production to enhance higher-twist processes, and point that the use of isolated hadrons as a signal for new physics at colliders can be affected by the presence of direct hadron production processes.}

\section{Introduction}

The most important discriminant of the twist of a perturbative QCD  subprocess in a hard hadronic  collision is the scaling of the inclusive invariant cross section~\cite{Brodsky:1973kr},
\begin{equation}\label{eq:scaling}
\sigma^{\rm inv} \equiv E\ \frac{\dd\sigma}{\dd^3 p}(A\ B\  \to C\ \X)  = \frac{F(\xt, \thetacm)}{\pt^n},
\end{equation}
at fixed $\xt = {2\pt/ \sqrts}$ and center-of-mass (CM) angle $\thetacm$.  In the original parton model the power fall-off is simply  $n=4$ since the underlying $2 \to 2$ subprocess amplitude for point-like partons is scale invariant, and there is no dimensionful parameter as in a conformal theory.
However, in general additional higher-twist (HT) contributions involving a larger number of elementary fields contributing to the hard subprocess, $\nactive>4$,  are also expected.   For example, the detected hadron $C$ can be produced directly in the hard subprocess as in an exclusive reaction. Unlike quark or gluon fragmentation, the direct processes do not waste same-side energy, thus involving minimal values of the momentum fractions $x_1$ and $x_2$ where parton distributions are maximal. Neglecting scaling violations in QCD, the scaling exponent $n$ is given by $n=2\nactive-4$.

The idea of direct hadron production was considered in the 1970's  to explain the large fixed $\xt$ scaling exponents reported at ISR  and fixed target FNAL energies~\cite{Brodsky:1973kr}.
However, there has been no comprehensive and quantitative analysis of the data up to now which could bring compelling evidence for such higher-twist effects. In these proceedings, we summarize the novel aspects discussed in our recent analysis~\cite{Arleo:2009ch}, namely:
\begin{itemize}
\item[(i)] a dedicated analysis of the most recent FNAL, RHIC and Tevatron data on large-$\pt$ hadrons, prompt photons and jets;
\item[(ii)] the systematic comparison of the experimental scaling exponents with NLO QCD expectations;
\item[(iii)] predictions for the top RHIC energy  and at the LHC.
\end{itemize}

\section{Analysis}

The exponent $\nnlo$ of mid-rapidity particle production has been computed in QCD at next-to-leading order (NLO) accuracy from Ref.~\cite{Aurenche:1998gv}. The $\xt$-dependence of $\nnlo$ at fixed $\pt$ has been determined for various hadron species ($\pi$, $K$, $p/\bar{p}$). At $\pt=10$~GeV the exponents increase slowly from $\nnlo\simeq5$ at small values of $\xt$ ($\xt=10^{-2}$) up to $\nnlo\simeq6$ at $\xt=0.5$ with almost no dependence on the specific hadron species. Remarkably, the exponent extracted in the prompt photon channel is below those of hadrons, by roughly one unit, close to the conformal limit, $n=4$, at the smallest values of $\xt$. This observation is understood from the (relative) absence of fragmentation processes and one less power in $\alpha_s$, leading to {less scaling violation} in this channel.

On the experimental side, the exponent $\nexp$ has been systematically extracted from measurements in hadronic collisions, from fixed-target to collider experiments. It is deduced from the comparison of $\xt$-spectra at different CM energies,
\begin{equation}\label{eq:nexp}
\nexp(\xt) \equiv-\frac{\ln\left(\sigmainv(\xt,\sqrt{s_1})\big/\sigmainv(\xt,\sqrt{s_2}) \right)}{\ln\left( \sqrt{s_1}\big/\sqrt{s_2} \right)}   
\end{equation}
which is equivalent to (\ref{eq:scaling}) at fixed $\xt$. The data sets include $\pi^0$ measurements by the E706 at FNAL~\cite{Apanasevich:2002wt} 
and by the PHENIX collaboration at RHIC~\cite{Adare:2007dg}.
At higher energies, the measurements of charged hadrons (or charged tracks) in $p$--$\bar{p}$ collisions at 
$\sqrts=630,\ 1800$~GeV by CDF~\cite{Abe:1988yu}  
and $\sqrts=500,\ 900$~GeV by UA1~\cite{Albajar:1989an}  
are included in the analysis. Also considered are prompt photon~\cite{Acosta:2002ya} and jet~\cite{Abe:1992bk} data obtained 
by CDF and D0 at $\sqrts=546,\ 630,\ 1800$~GeV.

\begin{figure*}[t]
  \begin{center}
\begin{minipage}{5.5cm}
    \includegraphics[height=6.5cm]{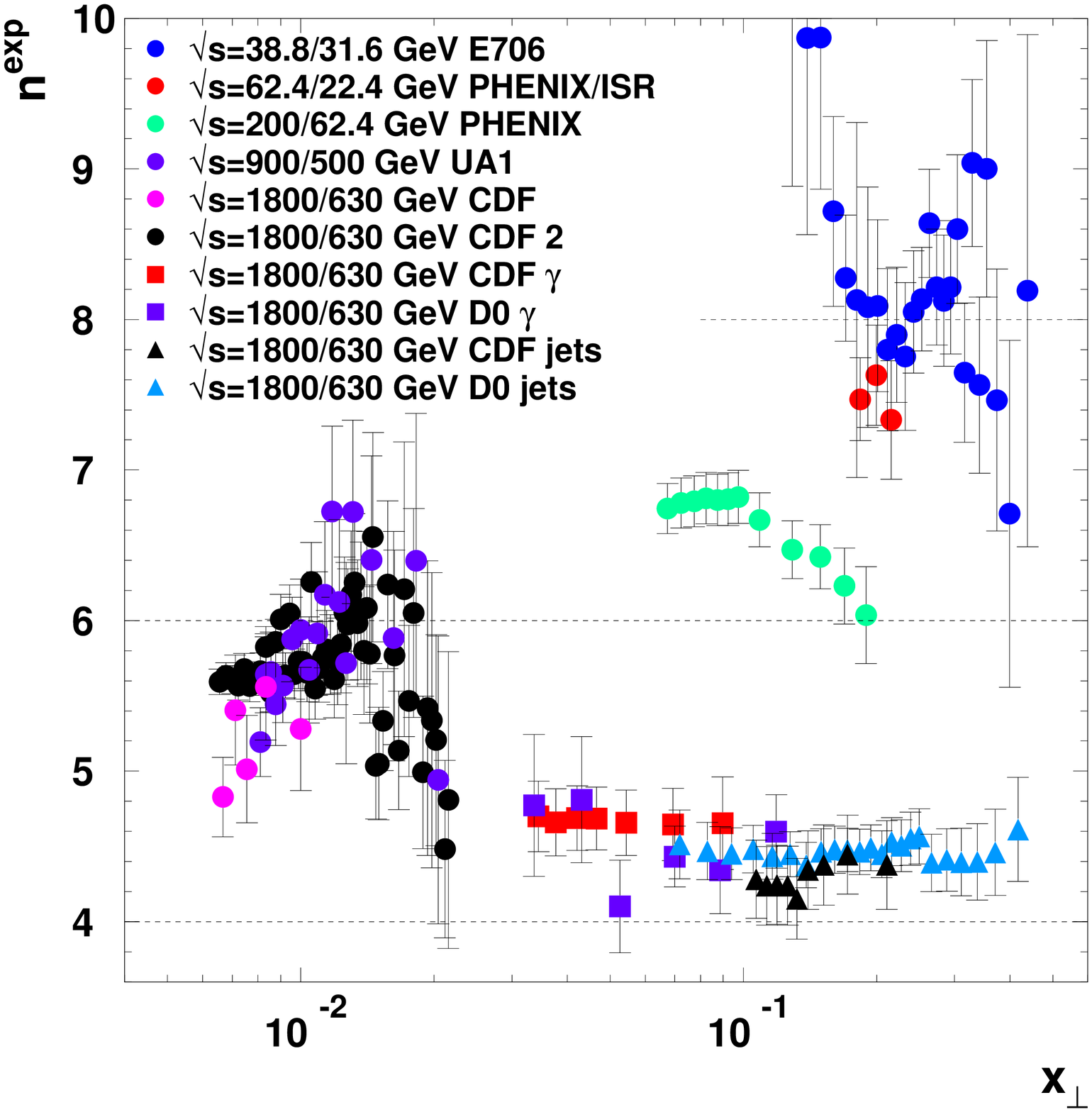}
\end{minipage}
~
\begin{minipage}{9.5cm}
    \includegraphics[width=10.cm,height=6.5cm]{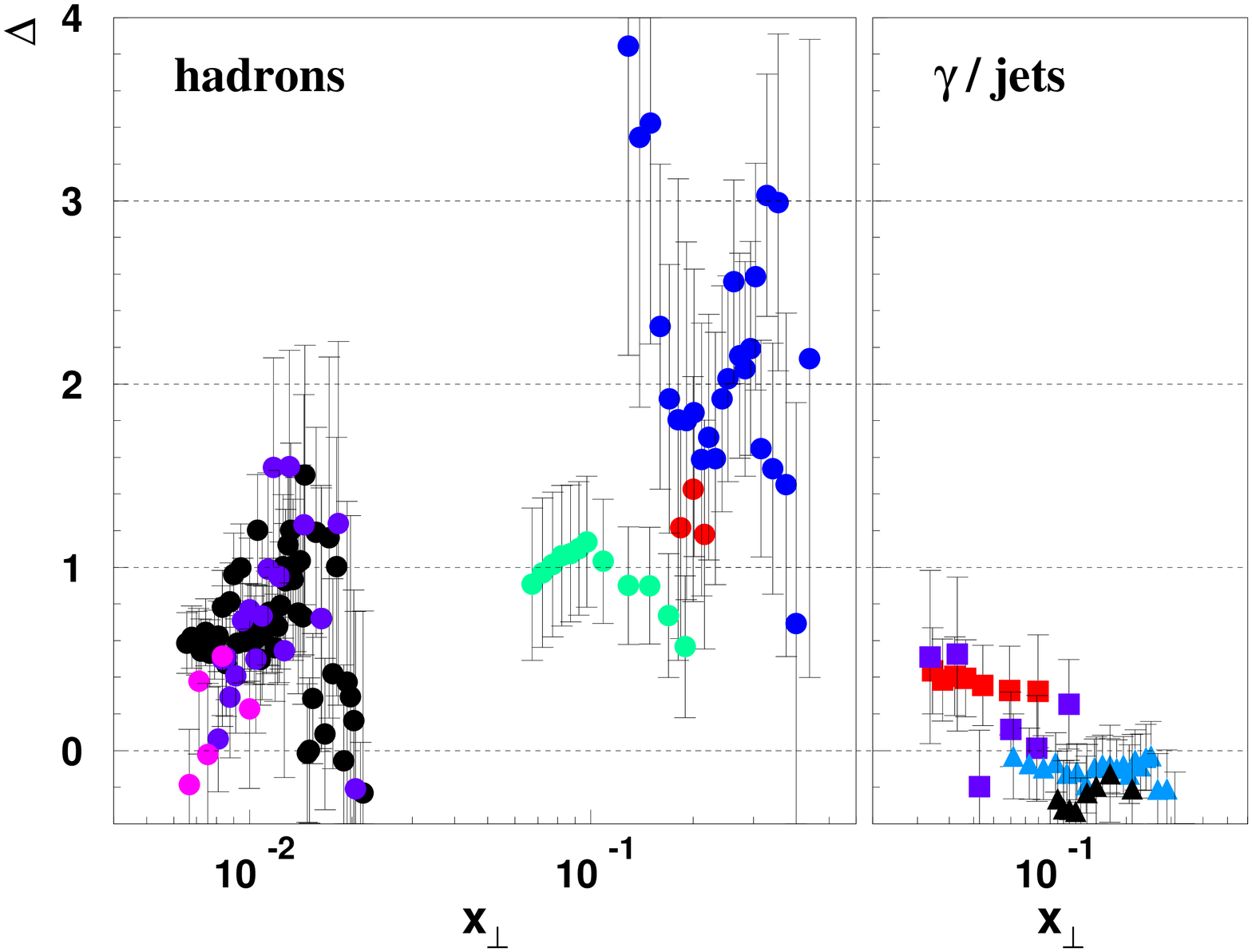}
\end{minipage}

\caption{{\it Left:} Values of $\nexp$ as a function of $\xt$ for $h^\pm/\pi^0$ (circles), $\gamma$ (squares) and jets (triangles). {\it Right: }$\Delta\equiv\nexp-\nnlo$ vs. $\xt$, error bars include the experimental and the theoretical uncertainties added in quadrature.}
  \label{fig:datatheory_compilation}
  \end{center}
\end{figure*}

\section{Results}

The hadron exponents plotted in Fig.~\ref{fig:datatheory_compilation} (left) exhibit a clear trend, with a significant rise of $\nexp$ as a function of $\xt$. Typical values of $\nexp$ are $\nexp\simeq5$--$6$ at small $\xt\simeq10^{-2}$ while PHENIX data 
point to a mean value $\nexp\simeq6.7$ at an intermediate $\xt\simeq10^{-1}$. At higher values of $\xt$, the measurements by PHENIX and E706 reveal an exponent even larger, $\nexp\simeq8$, confirming the results reported long ago at the ISR.
The exponents obtained in the photon and jet channels are strikingly different, showing almost no dependence on $\xt$. Importantly enough, the values obtained lie only slightly above the conformal limit, $\nexpga\simeq4.6$ and $\nexpjets\simeq4.4$, i.e. several units smaller than the exponents observed for hadrons.

In order to compare properly data and theory, the \emph{difference} between experimental and theoretical exponents, $\Delta(\xt)\equiv\nexp-\nnlo$, is plotted in the right panel of Fig.~\ref{fig:datatheory_compilation} for hadrons and photons/jets.  Note that the error bars include both experimental \emph{as well as} theoretical errors, added in quadrature. The theoretical uncertainty is estimated from the variation of renormalization/factorization scales from $\pt /2$ to $2\pt$. Fig.~\ref{fig:datatheory_compilation} (right) indicates that the hadronic exponents extracted experimentally prove significantly above the leading-twist (LT) predictions.
The discrepancy is moderate at small $\xt\sim10^{-2}$, $\Delta\simeq0.5$, but becomes increasingly larger at higher values of $\xt$: $\Delta \simeq 1$ at $\xt=10^{-1}$ and up to $\Delta\simeq2$ in the largest $\xt$ region.  In contrast, the scaling behavior observed for photons and jets are in very good agreement with the NLO predictions ($\Delta\simeq0$).

\section{Discussion}

Part of the discrepancy reported in hadron production data at large $\xt\sim1$ could occur  because of the appearance of large threshold logarithms, $\ln(1-\xt)$, which should be resummed to all orders in perturbation theory~\cite{Laenen:2000ij}. It would therefore be most interesting to investigate whether or not threshold resummation might bring data and theory in agreement. Note however that the discrepancy is also observed at small values of $\xt\sim10^{-2}$, where such effects are usually expected to be small.

A natural explanation for the large exponents observed in the hadron channel is the presence of important HT contributions from processes in which the detected hadron is produced directly in the hard subprocess, because of the dimension of the hadron distribution amplitude.
  In contrast, particles having no hadronic structure like isolated photons and jets are much less sensitive to such HT contributions and should behave closer to LT expectations, as observed. Another piece of evidence for HT effects is the larger exponents for protons than for pions observed at the ISR. As discussed in~\cite{Arleo:2009ch}, the difference between the {\it direct} proton and pion scaling exponent is $n_{p}-n_{\pi}=2$ ($n_{p}=8$, $n_{\pi}=6$) instead of $n_{p}-n_{\pi}\simeq0$ at LT. 
  The experimental value obtained from the ISR, $n_{p}-n_{\pi}\simeq1$, thus reflects the mixture of LT and HT contributions to the total cross section. It has also been noted~\cite{Brodsky:2008qp} that the presence of color-transparent  HT subprocesses can account for anomalous features of proton production in heavy ion collisions~\cite{Adler:2003kg}.

\begin{figure}[t]
\begin{center}
\includegraphics[width=11.6cm,height=5.4cm]{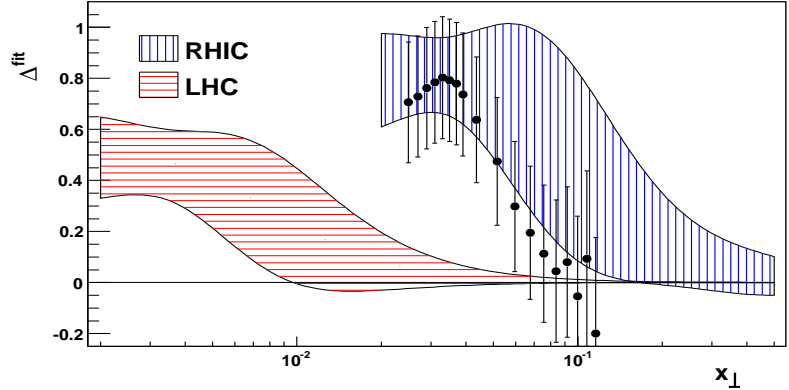}
\end{center}
\caption{Predicted difference between the experimental and NLO scaling 
exponent at RHIC ($\sqrt{s}=200,500 $ GeV) and the LHC  ($\sqrt{s}=7$~TeV as compared to $\sqrt{s}=1.8$~TeV), compared to PHENIX preliminary measurements.}
\label{fig:lhc}
\end{figure}

Finally, we discuss the phenomenological consequences of possible HT contributions to hadron production in $p$--$p$ collisions at RHIC and LHC. In order to obtain qualitative predictions, the difference $\Delta$ between the experimental and the NLO exponent has been fitted to the hadron data currently analyzed. The typical values of $\Deltafit$ expected at RHIC (taking $\sqrts=200, 500$~GeV) and at LHC  ($\sqrts=7$~TeV, compared to $\sqrts=1.8$~TeV at Tevatron) are plotted as a function of $\xt$ in Fig.~\ref{fig:lhc}. At RHIC, $\Deltafit$ is slightly below 1 at small $\xt\lesssim5.10^{-2}$ but decreases towards zero at larger $\xt$. The predictions turn out to be in very good agreement, both in shape and magnitude, with the PHENIX preliminary measurements~\cite{bazilevsky} performed at $\sqrt{s}=500$~GeV. At LHC, smaller deviations with NLO expectations are expected because of the large values of $\pt$ probed at high energy: $\Deltafit\simeq0.5$  below $\xt=5\times10^{-3}$ and smaller above.
%
%
In order to enhance the HT contribution to hadron production, we suggest to trigger on {\it isolated} hadrons, i.e. with small hadronic activity in their vicinity. We also point that the use of isolation cuts, usually applied for prompt photons, will strongly suppress LT processes. Consequently, the scaling exponents of isolated hadrons are expected to be somewhat larger than those in the inclusive channel. The use of isolated hadrons as a signal for Higgs production or new physics scenarios~\cite{nisatisirois} might be confused by the presence of direct hadron production processes.

We thank A. Bazilevsky for providing us with Fig.~\ref{fig:lhc}. FA thanks CERN-TH for hospitality. SJB was supported by the Department of Energy under contract DE-AC02-76SF00515.  DSH was supported by the International Cooperation 
Program of the KICOS and the Korea Research Foundation Grant (KRF-2008-313-C00166).  AMS was supported by the Department of Energy
under contract DE-AC02-98CH10886.

\providecommand{\href}[2]{#2}\begingroup\raggedright\endgroup

\end{document}